# Boundary-induced topological chiral extended states in Weyl metamaterial waveguides

Ning Han[1,2,3,4,9], Fujia Chen[1,2,3,4,9], Mingzhu Li[5], Rui Zhao[1,2,3,4], Wenhao Li[1,2,3,4], Qiaolu Chen[1,2,3,4], Li Zhang[1,2,3,4], Yuang Pan[1,2,3,4], Jingwen Ma[6], Zhi-Ming Yu[7,8], Hongsheng Chen[1,2,3,4†] and Yihao Yang[1,2,3,4†]

1. *Interdisciplinary Center for Quantum Information, State Key Laboratory of Extreme Photonics and Instrumentation, ZJU-Hangzhou Global Scientific and Technological Innovation Center, Zhejiang University, Hangzhou 310027, China.*

2. *International Joint Innovation Center, The Electromagnetics Academy at Zhejiang University, Zhejiang University, Haining 314400, China.*

3. *Key Lab. of Advanced Micro/Nano Electronic Devices & Smart Systems of Zhejiang, Jinhua Institute of Zhejiang University, Zhejiang University, Jinhua 321099, China.*

4. *Shaoxing Institute of Zhejiang University, Zhejiang University, Shaoxing 312000, China.*

5. *School of Information and Electrical Engineering, Hangzhou City University, Hangzhou 310015, China.*

6. *Department of Electronic Engineering, The Chinese University of Hong Kong, Shatin, New Territories, Hong Kong.*

7. *Centre for Quantum Physics, Key Laboratory of Advanced Optoelectronic Quantum Architecture and Measurement (MOE), School of Physics, Beijing Institute of Technology, Beijing 100081, China.*

8. *Beijing Key Laboratory of Nanophotonics and Ultrafine Optoelectronic Systems, School of Physics, Beijing Institute of Technology, Beijing 100081, China.*

9. *These authors contributed equally: Ning Han, Fujia Chen.*

†*E-mail: yangyihao@zju.edu.cn (Y. Y.) ; hansomchen@zju.edu.cn (H. C.).*

**In topological physics, it is commonly understood that the existence of the boundary states of a topological system is inherently dictated by its bulk. A classic example is that the surface Fermi-arc states of a Weyl system are determined by the chiral charges of Weyl points within the bulk. Contrasting with this established perspective, here, we theoretically and experimentally discover a family of topological chiral bulk states extending over photonic Weyl metamaterial waveguides, solely induced by the waveguide boundaries, independently of the waveguide width. Notably, these bulk states showcase discrete momenta and function as "wormhole" tunnels that connect Fermi-arc surface states living in different two-dimensional (2D) spaces via a third dimension. Our work offers a magnetic-field-free mechanism for robust chiral bulk transport of waves and highlights the boundaries as a new degree of freedom to regulate bulk Weyl quasiparticles.**

Weyl semimetals and their classical-wave analogues epitomise a captivating category of topological materials with band structures showcasing two-fold linear crossings in three-dimensional (3D) momentum space[1-12]. These crossing points, also known as Weyl points, are monopoles of Berry flux characterised by nonzero chiral charges or Chern numbers. Owing to the Weyl points within the bulk, there exist 2D open Fermi-arc states on the surfaces, which connect the projections of Weyl points with opposite chiral charges[13-22]. Depending on the distributions and the numbers of the Weyl points in the bulk, the surface Fermi-arc states can exhibit various configurations.

Here, we theoretically and experimentally study photonic Weyl metamaterial waveguides, Weyl metamaterials truncated by a pair of boundaries, and uncover a class of chiral bulk states extending over the waveguides, induced solely by the boundaries, independently of the waveguide width. These chiral extended states (CESs) are somewhat counterintuitive, as they are bulk states but regulated by the boundaries, which are different from the traditional physical picture where the boundary states are determined by the bulk properties. Also fundamentally different from the magnetic-field-induced chiral Landau level bulk states[23-27], the CESs do not require (artificial) magnetic fields, which greatly facilitates their practical applications. Moreover, the CESs have a topological origin associated with the chiral charges of the Weyl points, which provides a new way to probe the chiral charges of Weyl points in a finite-size system. Via the first-principles studies and the direct pump-probe measurements, we surprisingly unveil that the topological CESs showcase discrete momenta and serve as "wormhole" tunnels that bridge the Fermi-arc surface states residing in different 2D spaces via a third dimension. Finally, the pseudospin-locking and discrete-momentum nature of the CESs enables significant suppression of the scattering losses caused by obstacles.

To begin with, we consider a 3D photonic Weyl medium (see Fig. 1a) described by an effective Hamiltonian,

$$\hat{H} = \hat{\sigma}_x \left( k_x^2 - \Delta^2 \right) v_x + \hat{\sigma}_y \left( k_y^2 - \Delta^2 \right) v_y + \hat{\sigma}_z k_z v_z, \qquad (1)$$

where $\hat{\sigma}_{x(y,z)}$ represent the Pauli matrices, $\Delta$ is a shift of the Weyl points from the origin of momentum space, and $v_{x(y,z)}$ are Weyl velocities, respectively. The Weyl medium hosts four ideal Weyl points at $(-\Delta, -\Delta)$, $(-\Delta, +\Delta)$, $(+\Delta, -\Delta)$, $(+\Delta, +\Delta)$, and carries a chiral charge of $+1, -1, -1, +1$, respectively. It can be readily realised using metallic or dielectric photonic metamaterials[14], as we will show later.

Now, we truncate the Weyl medium at $z = 0$ and $z = d$, with boundaries enforcing the derivatives of the wavefunctions to be zero. At the boundaries, the eigenstates $\left| \psi \left( k_x, k_y, z \right) \right\rangle$ propagating in the Weyl waveguide satisfy,

$$\hat{M} \left| \psi \left( k_x, k_y, z_0 \right) \right\rangle = \left| \psi \left( k_x, k_y, z_0 \right) \right\rangle, \qquad (2)$$

where $z_0 = 0$ $(z_0 = d)$, $\hat{M}(\theta) = \hat{\sigma}_x \sin(\theta) + \hat{\sigma}_y \cos(\theta)$, with $\theta$ being phenomenologically determined[28-30]. The eigenstates $\left| \psi \left( k_x, k_y, z \right) \right\rangle$ are determined by both the bulk Hamiltonian $\hat{H}$ and the boundary Hamiltonian $\hat{M}$.

In the Weyl waveguide, many eigenstates exist. Each eigenstate is a superposition of two linearly independent wavefunctions $\left| \psi \left( k_x, k_y, z \right) \right\rangle \propto \left[ \alpha_1 e^{ipz} + \alpha_2 e^{-ip(z-d)} \right]$, where $\alpha_1$ and $\alpha_2$ are two coefficients, and $p$ is the wavevector (see details in Supplementary Note 2). When $p$ is a real (imaginary) number, the eigenstates are bulk (surface) states. Substituting the wavefunction $\left| \psi \right\rangle$ to Eqs. (1)-(2), and setting $\theta = \pi/4$ (which matches well with our realistic design as we discuss later), we obtain the guided bulk modes with energy bands, $\varepsilon_m = \pm \sqrt{ \left( m\pi \right)^2 + d^2 \left( k_x^2 - \Delta^2 \right)^2 + d^2 \left( k_y^2 - \Delta^2 \right)^2 } \Big/ d$, where $m \geq 1$ is the order of the guided modes. The resulting energy bands are shown in Fig. 1b, where a pseudo-bandgap opens between two first-order guided bulk modes, with a bandwidth $\delta\varepsilon = 2\pi/d$ determined by $d$.

Also, we can obtain the surface states at each $(k_x, k_y)$, $\left|\psi_{surface}\left(k_x, k_y, z\right)\right\rangle = \beta_1\left|\psi_1\left(k_x, k_y, z\right)\right\rangle + \beta_2\left|\psi_2\left(k_x, k_y, z\right)\right\rangle$, where $\beta_1$ and $\beta_2$ are two coefficients, and $\left|\psi_n\left(k_x, k_y, z\right)\right\rangle \propto e^{\pm ipz}$ ($n=1, 2$) represent surface-wave-like eigenstates residing at opposite boundaries, with a decay constant (see detail in Supplementary Note 2),

$$|p| = \left|k_x^2 - k_y^2\right|/\sqrt{2} . \tag{3}$$

The decay constant $|p|$ in Eq. (3) is a key parameter. As shown in Fig. 1c, when $|k_x| = |k_y| = k$, $|p| = 0$, corresponding to the bulk modes extending in the entire Weyl medium (see the white dashed lines). The corresponding eigenvalues $\varepsilon_{CB}$ and eigenstates $\left|\psi_{CB}\right\rangle$ of the extended modes are,

$$\varepsilon_{CB} = \sqrt{2}\left(k^2 - \Delta^2\right), \quad \left|\psi_{CB}\right\rangle = \begin{pmatrix} (1-i)/\sqrt{2} \\ 1 \end{pmatrix}, \tag{4}$$

respectively. There are four extended modes for a given eigenvalue $\varepsilon_{CB}$; each extended mode derives from a Weyl point (see Figs. 1b and 1c). Interestingly, around a Weyl point in the case without boundaries, the extended states with opposite propagation directions are always paired; after truncation, however, the extended states locked with pseudospins can only propagate chirally, either in the left or right direction, dubbed as CESs. Notably, distinct from other surface states, the CESs are a special solution that is independent of the waveguide width $d$.

Physically, considering the Weyl point around $(+\Delta, +\Delta)$ in the case without the boundaries, there exists a pair of counterpropagating eigenstates with even or odd symmetry under the operator $\hat{P} = \left(\hat{\sigma}_x + \hat{\sigma}_y\right)/\sqrt{2}$, respectively, along $k_x = k_y$; after applying the boundary conditions to form a Weyl waveguide, only the forward-propagating extended state with even symmetry survives (see Fig. 1d and more details in Supplementary Notes 1-3). Therefore, the CESs are induced by the boundary effect.

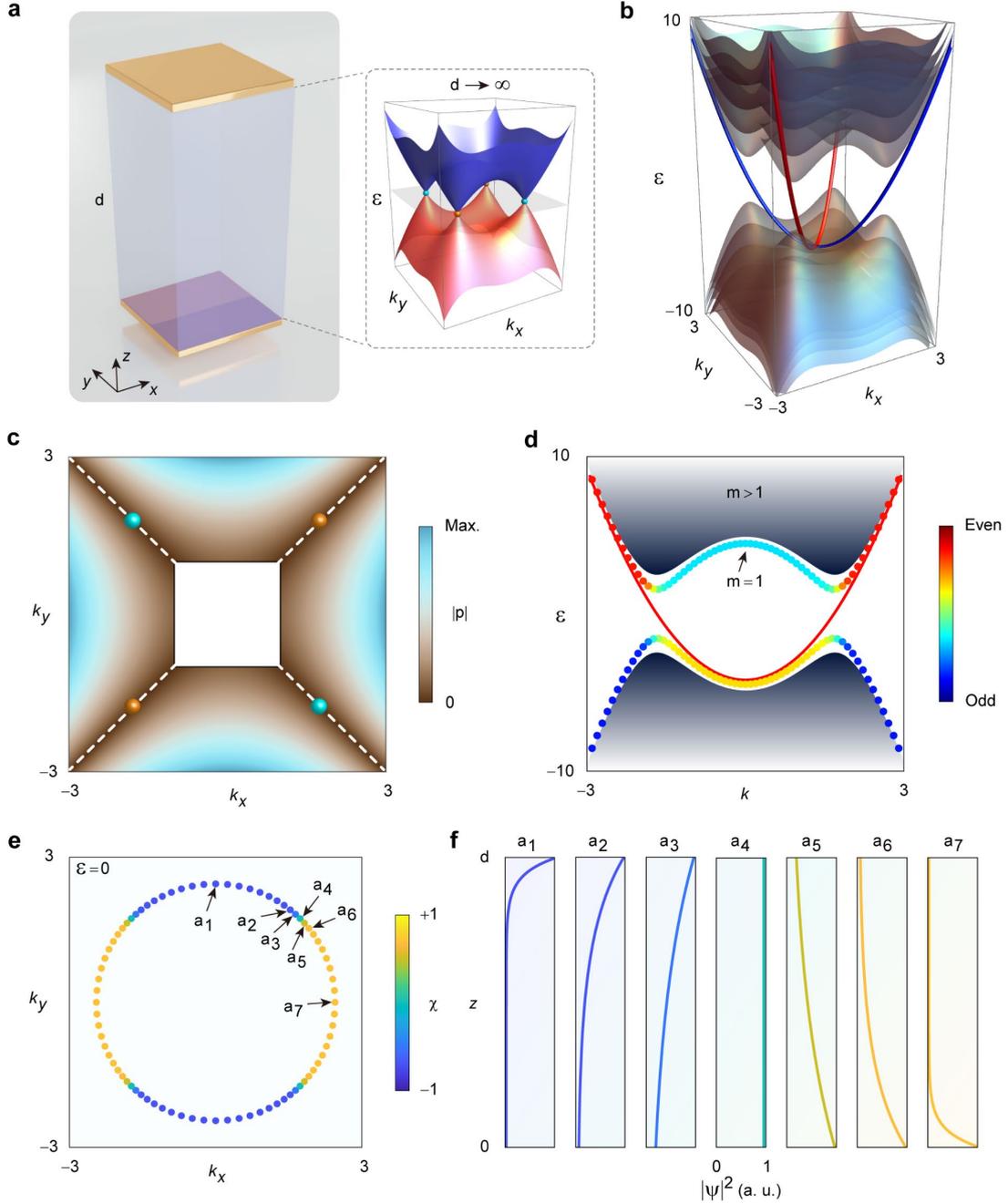

**Fig. 1. Theoretical model of the CESs in a Weyl metamaterial waveguide. a** Schematic illustration of the Weyl waveguide. The yellow planes represent the truncation boundary. The inset shows the Weyl cones of the middle medium ($d \to \infty$ and $\Delta^2 = 3$). **b** Band structures of the bulk modes ($d=2$, $\theta = \pi/4$, and $\Delta^2 = 3$). The red and blue lines denote the CESs. **c** Distribution of attenuation constant $p$ in momentum space. The orange and cyan dots denote Weyl points with positive and negative topological charges, respectively. **d** Parity of the eigenmodes along $k_x = k_y$. The CESs have even symmetry. The other bulk modes are hybridised. **e** 2D isofrequency contour and $\chi$ of the eigenmodes at $\varepsilon = 0$. **f** Density distributions of the eigenmodes indicated in **e**.

When $p \neq 0$, the eigenstates $\left| \psi_{surface} \left( k_x, k_y, z \right) \right\rangle$ are surface-wave-like states bound to either boundary ($z = 0$ or $z = d$ in Fig. 1a). We define a dimensionless factor $\chi$ to characterise the asymmetric distribution of the eigenstates $\chi = (A - B)/(A + B)$, where $A = \left\| \psi(k_x, k_y, 0) \right\rangle \right|^2$ and $B = \left\| \psi(k_x, k_y, d) \right\rangle \right|^2$ represent density distributions at the bottom and top surfaces of the Weyl waveguide, respectively. Intuitively, $\chi = -1$, $\chi = 0$, and $\chi = 1$ represent the top surface states, the CESs, and the bottom surface states, respectively.

At $\varepsilon = 0$, we calculate $\chi$ along the 2D isofrequency contour, as shown in Fig. 1e (see Supplementary Note 2 for the detailed discussion). One can see that $\chi = -1$ ($\chi = 1$) dots form two open arcs, corresponding to the top (bottom) Fermi-arc surface states and connecting to the projection of two oppositely charged Weyl points. At the junction between the top and bottom Fermi-arc surface states, $\chi$ dramatically charges from -1 to +1, which must pass through $\chi = 0$, corresponding to the CESs (see Fig. 1f). Thus, the CESs serve as tunnels connecting the Fermi-arc surface states living in top and bottom 2D spaces via a third dimension. Moreover, it can be proved that the CES has a topological origin associated with the chiral charges of Weyl points (see details in Supplementary Note 4 and Note 5).

As a platform for the experimental realisation, we choose the saddle-shaped Weyl metamaterial sandwiched between copper plates. This Wely metamaterial is known for hosting ideal Weyl points that isolate from other trivial bands[14]. It has the following structural parameters: $a_x = 6$ mm, $a_y = 6$ mm, $a_z = 6$ mm, $h = 4$ mm, $w = 0.25$ mm, $L = 3.6$ mm, $r = 0.4$ mm, and $R = 0.7$ mm; the thickness of the copper film is 0.035 mm; and it is embedded in a dielectric background with a relative permittivity 2.6 (see Fig. 2a). The wave phenomenon of this 3D Weyl system around the Weyl frequency of 7.8 GHz can be well captured by the Hamiltonian Eq. (1).

Then, we perform first-principles studies of the Weyl metamaterial waveguide. From the band structures in Fig. 2b, one can see that around the Weyl frequency, a pseudo-bandgap opens, in which only a single gapless mode exists at each $k$-point, as indicated by the purple surface. In the pseudo-bandgap, the eigenmodes along the isofrequency contour (at 8.1 GHz) show the transition from the top Fermi-arc surface states to CESs and to bottom Fermi-arc surface states, as expected (see Fig. 2c). To quantitatively analyse the eigenmode distribution, we define a dimensionless factor $\delta$ similar to $\chi$, $\delta = \left( \sum_{Bottom} |E| - \sum_{Top} |E| \right) \Big/ \left( \sum_{Bottom} |E| + \sum_{Top} |E| \right)$, where $\sum_{Bottom} |E|$ and $\sum_{Top} |E|$ represent the total eigen electric fields $|E|$ along a surface (indicated in Fig. 2c) near the bottom and top of the Weyl waveguide, respectively. We indeed see a transition of $\delta$ from -1, to 0, and to +1 in the first quadrant of the isofrequency contour (see Fig. 2d), matching well with our theory. By checking the mode profile of each eigenmode, we find four CESs along $\Gamma M$ direction, corresponding to four Weyl points (see Fig. 2e).

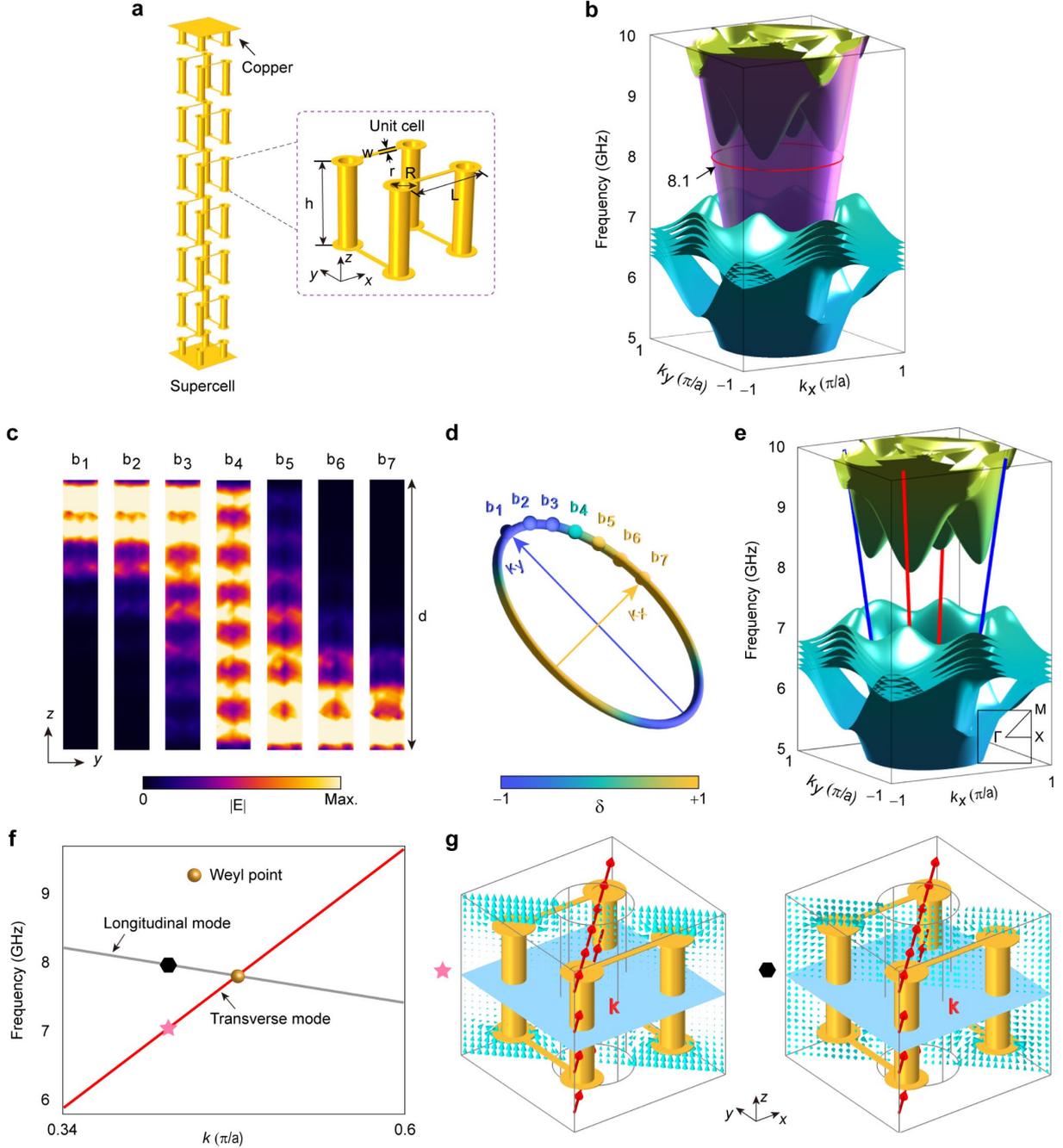

**Fig. 2. First-principles study. a** Supercell of the Weyl metamaterial waveguide. **b** Band structure of the Weyl waveguide. The purple surface corresponds to the eigenstates in the pseudo-bandgap. **c** Eigen electric field $|E|$ profiles at 8.1 GHz. **d** 2D isofrequency contour of the eigenmodes indicated in **c**. $\delta$ is used to quantitatively analyse the eigenmode distribution. **e** Dispersion of the CESs. The surface states are removed for clarity. **f** Transverse mode (TM) and longitudinal mode (LM) near the Weyl frequency. **g**

Eigen electric field profiles of the TM and LM, corresponding to the pink pentagram and black hexagon in **f**, respectively. The red **k** represents the wavevector.

The above phenomenon can be further understood from the perspective of electromagnetic theory. It is known that the present Weyl metamaterial can be described as a homogenous medium based on the effective medium theory (see Supplementary Note 6 and Note 8 for a detailed discussion). Along the $\Gamma M$ direction, it supports two fundamental modes, transverse mode (TM) and longitudinal mode (LM); the former (latter) has electric components vertical to (parallel with) the wavevector. In our realistic structure, one can also find a TM and LM (see Figs. 2f and 2g). Particularly, $E_z$ is dominant and confined in the gap between metallic structures in the TM, while $E_x$ and $E_y$ are dominant and distribute uniformly in the LM. When truncating the Weyl metamaterial with a perfect electric conductor (PEC) boundary at the middle plane of the metallic structure (see the blue planes in Fig. 2g), the $E_x$ and $E_y$ components parallel to the PEC boundary are not allowed. This leads to the suppression of LM and the survival of TM whose $E_x$ and $E_y$ components are nearly zero in the middle plane. The other modes on the Weyl cone can be viewed as a superposition of LM and TM, and are also suppressed by the PEC boundary. In terms of symmetry, TM (LM) has even (odd) symmetry, which agrees with our theoretical analysis.

Next, we perform experiments to demonstrate the CESs in the Weyl metamaterial waveguide, with the fabricated sample shown in Fig. 3a. In the experiment, a point-like source is vertically inserted into the central position of the sample through the air holes of the Weyl waveguide to excite the CESs (see Fig. 3b and more field distributions in Supplementary Note 9). A second electric dipole antenna is inserted into the middle layer of the sample from the top to probe the field distributions of the CESs. The field distributions in the Weyl metamaterial waveguide are measured point by point with a spatial resolution of 6 mm, allowing for detailed analysis. By applying a 2D spatial Fourier transform to the complex field patterns measured in real space at different frequencies, we can obtain the band structure of the measured modes.

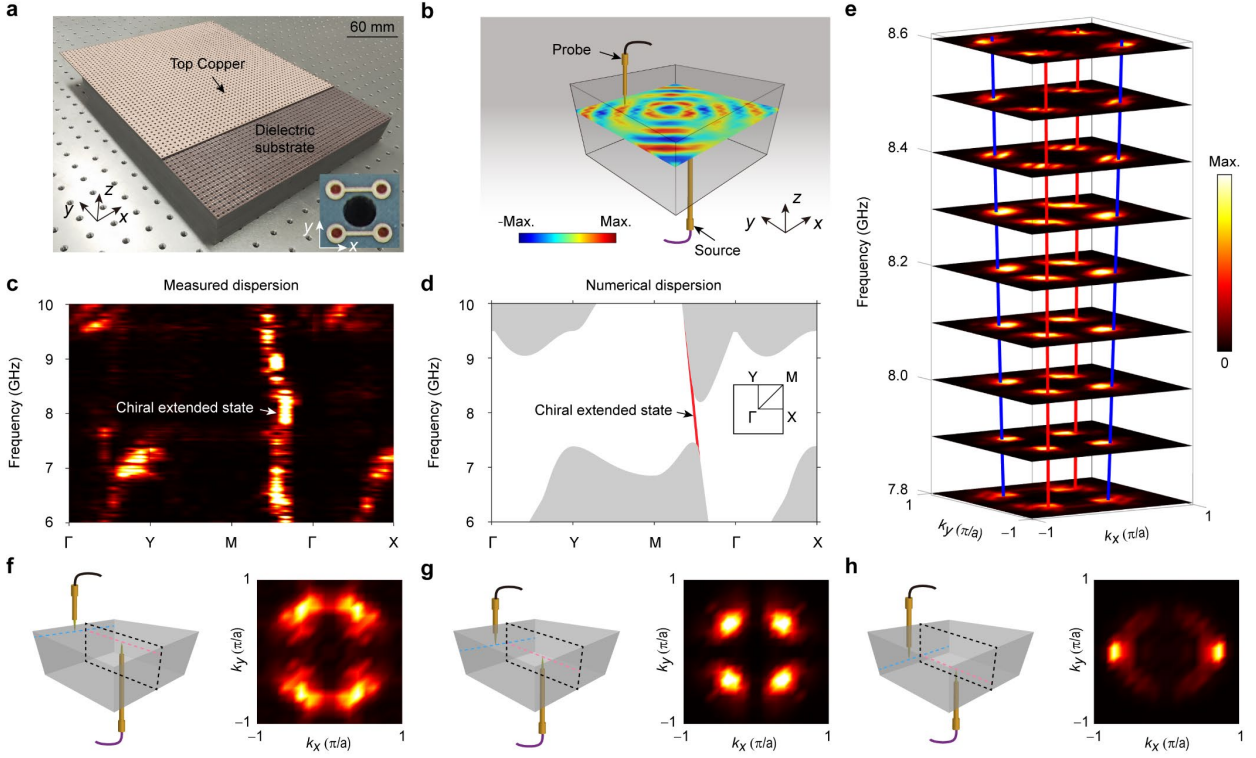

**Fig. 3. Observation of the CESs and the "wormhole" tunnel effect. a** Photograph of the fabricated Weyl waveguide. The sample has $50 \times 50 \times 7$ unit cells. The inset shows a unit cell. For improved visibility, a section of the top copper cover plate has been displaced. **b** Experimental setup for the CES measurements. The electric field distribution is measured in the middle plane of the sample at 8.1 GHz. **c, d** Measured and numerical dispersion of the CESs. The grey regions represent the band structure of the Weyl metamaterial waveguide. **e** Measured 2D isofrequency contours of the CESs at different frequencies. Red and blue lines guide the eye. **f-h** Measured 2D isofrequency contours of the top Fermi-arc states, the CES, and the bottom Fermi-arc states at 8.1 GHz. The color bar represents the electric field intensity $|E_z|$. The left panel in each figure shows the corresponding experimental setup.

The measured results are shown in Fig. 3c. It is evident from the measured band structure in Fig. 3c that the CESs exist within a specific frequency range in the Weyl metamaterial waveguide, which is highly consistent with the numerically simulated counterpart (see Fig. 3d). We also plot the measured 2D isofrequency contours of the CESs as a function of frequency (see Fig. 3e). Indeed, we can find four discretised CESs at each frequency within the pseudo-bandgap, each derived from a Weyl point. Our results, thus, provide direct experimental evidence for the existence of CESs in Weyl metamaterial waveguides. Interestingly, in a conventional parallel waveguide with a multiple-wavelength thickness (e.g., two wavelengths in our case), there are always numerous high-order eigenmodes. The eigenmodes are usually more complicated when filling the parallel waveguide with resonant metamaterials. However, only a single mode exists in the pseudo-bandgap, which is counterintuitive from the perspective of the traditional electromagnetic theory.

The CESs serve as the "wormhole" tunnels, connecting the Fermi-arc surface states living in top and bottom 2D space via a third dimension. To confirm this point, we perform additional measurements, where the source and probe are placed at the top (bottom) layer to measure the top (bottom) Fermi-arc-like surface states of the Weyl waveguide. In Figs. 3f-h, we plot the measured surface-state isofrequency contours together with the measured bulk-state counterpart at 8.1 GHz. One can see that the CESs are at the joint point between the top and bottom Fermi arcs, confirming the CESs as the "wormhole" tunnels.

The CESs exhibit suppressed scattering when encountering obstacles (see Figs. 4a and 4b). This is because the CESs are locked to pseudospins and rarely scatter to each other due to the discretised momentum and the large separation between them in momentum space. Also, they rarely scatter to the surface states due to the near orthogonality between the surface states and the CESs (see Supplementary Note 10). The scattering-suppressed propagation of CESs is fundamentally different from the propagation of states in conventional media with circle-like isofrequency contours. The latter can be easily scattered to other states with matched momenta when encountering obstacles (see Fig. 4d).

To demonstrate the above phenomenon, we design an experiment where a large 3D obstacle (a metallic brick with an orientation angle of 30°) with the size of $5 \times 5 \times 1$ unit cells is placed at the centre of the sample (see Fig. 4e). From the measured field distribution in Figs. 4f (4j) (without obstacle) and Figs. 4g (4k) (with an obstacle), we can see that the field patterns between the cases with or without obstacles are very close, indicating a tiny scattering from the obstacle. This is more evident from the momentum-space results, where only the CESs in the first quadrant are excited, and the scattering to other bulk states is weak. As a comparison, we also show the field patterns and the momentum-space energy distribution for the case outside the pseudo-bandgap (at 9.3 GHz). As shown in Figs. 4h (4l) (without obstacle) and Figs. 4i (4m) with an obstacle), we can see a strong scattering caused by the obstacle.

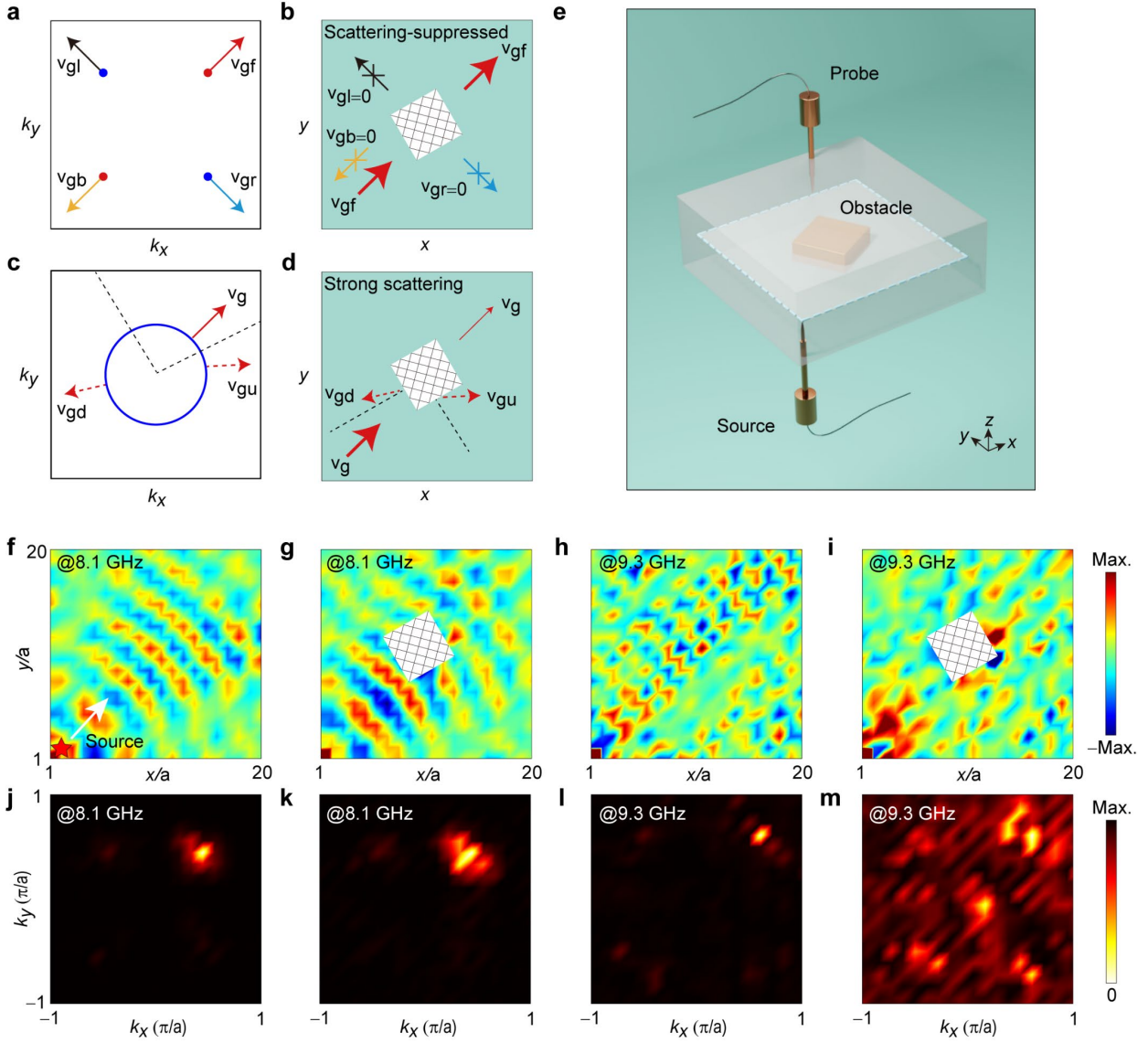

**Fig. 4. Scattering-suppressed propagation of the CESs. a-d** Schematics of scattering process when the CESs (**a, b**; discrete momenta: red and blue dots) or a normal state with circle-like isofrequency contours (**c, d**; blue line) encounters a large obstacle (shaded areas with the orientation angle of 30°). **a, c** Momemtum-space analyse. **b, d** Real-space analysis. Red dashed arrows in **c, d** indicate scattering with group velocities $v_{gu}$ and $v_{gd}$, respectively. **e** Experimental setup. The field pattern is measured in the middle plane (white plane) of the sample. **f-i** Electric field distributions are measured at 8.1 GHz (CESs) and 9.3 GHz (trivial bulk states). The red pentagram represents the dipole source. **j-m** Isofrequency contours $|E_z|$ of the CESs (trivial bulk states) plotted in momentum space.

To summarise, we theoretically predict and experimentally discover the topological CESs in photonic Weyl metamaterial waveguides, induced by the boundary effect, independently of the waveguide width. Such CESs serve as "wormhole" tunnels that connect the top and bottom

Fermi-arc surface states via a third dimension and exhibit scattering-suppressed propagation. The underlying physics has been well understood from the effective Hamiltonian theory, the first-principles studies, and the electromagnetic wave theory. Our work thus provides a magnetic-field-free solution to robust chiral transport of electromagnetic waves in the bulk medium (as well as other waves; see Supplementary Note 11 and Note 12 for the detailed discussions) and reveals the boundaries as a new degree of freedom to regulate bulk Weyl quasiparticles for exploring novel physical phenomena related, such as chiral anomaly[23].

## References

1. Lu, L., Fu, L., Joannopoulos, J. D. & Soljačić, M. Weyl points and line nodes in gyroid photonic crystals. Nat. Photonics **7**, 294-299 (2013).

2. Lu, L. et al. Experimental observation of Weyl points. Science **349**, 622 (2015).

3. Xiao, M., Chen, W., He, W. & Chan C. T. Synthetic gauge flux and Weyl points in acoustic systems. Nat. Phys. **11**, 920 (2015).

4. He, H. et al. Topological negative refraction of surface acoustic waves in a Weyl phononic crystal. Nature **560**, 61 (2018).

5. Li, F., Huang, X., Lu, J., Ma, J. & Z. Liu. Weyl points and Fermi arcs in a chiral phononic crystal. Nat. Phys. **14**, 30 (2018).

6. Wang, D. et al. Photonic Weyl points due to broken time-reversal symmetry in magnetised semiconductor. Nat. Phys. **15**, 1150 (2019).

7. Yang, Y. et al. Topological triply degenerate point with double fermi arcs. Nat. Phys. **15**, 645 (2019).

8. Luo, L. et al. Observation of a phononic higher-order Weyl semimetal. Nat. Mater. **20**, 794 (2021).

9. Ilan, R., Grushin, A. G. & Pikulin, D. I. Pseudo-electromagnetic fields in 3D topological semimetals. Nat. Rev. Phys. **2**, 29 (2020).

10. Xu, S. et al. Discovery of a Weyl fermion semimetal and topological Fermi arcs. Science **349**, 613 (2015).

11. Hu, J., Xu, S., Ni, N. & Mao, Z. Transport of topological semimetals. Annu. Rev. Mater. Res. **49**, 207 (2019).

12. Xiao, M., Lin, Q. & Fan, S. Hyperbolic Weyl point in reciprocal chiral metamaterials. Phys. Rev. Lett. **117**, 057401 (2016).

13. Soluyanov, A. A. et al. Type-II Weyl semimetals. Nature **527**, 495 (2015).

14. Yang, B. et al. Ideal Weyl points and helicoid surface states in artificial photonic crystal structures. Science **359**, 1013 (2018).

15. Ma, S. et al. Linked Weyl surfaces and Weyl arcs in photonic metamaterials. Science **373**, 572 (2021).

16. Yang, Y. et al. Ideal unconventional Weyl point in chiral photonic metamaterial. Phys. Rev. Lett. **125**, 143001 (2020).

17. Fu, Y & Qin, H. Topological phases and bulk-edge correspondence of magnetised cold plasmas. Nat. Commun. **12**, 3924 (2021).

18. Liu, G. G. et al. Topological Chern vectors in three-dimensional photonic crystals. Nature **609**, 925-930 (2022).

19. Chen, Q. et al. Discovery of a maximally charged Weyl point. Nat. Commun. **13**, 7359 (2022).

20. Pan, Y. et al. Real higher-order Weyl photonic crystal. Nat. Commun. **14**, 6636 (2023).

21. Xi, X. et al. Topological antichiral surface states in a magnetic Weyl photonic crystal. Nat. Commun. **14**, 1991 (2023).

22. Wang, D., Jia, H., Yang, Q., Hu, J., Zhang, Z. Q. & Chan, C. T. Intrinsic triple degeneracy point bounded by nodal surfaces in chiral photonic crystal. Phys. Rev. Lett. **130**, 203802 (2023).

23. Pikulin, D. I., Chen, A. & Franz, M. Chiral anomaly from strain-induced gauge fields in Dirac and Weyl semimetals. Phys. Rev. X **6**, 041021 (2016).

24. Jia, H. et al. Observation of chiral zero mode in inhomogeneous three-dimensional Weyl metamaterials. Science **363**, 148-151 (2019).

25. Peri, V., Serra-Garcia, M., Ilan, R. & Huber, S. D. Axial-field-induced chiral channels in an acoustic Weyl system. Nat. Phys. **15**, 357-361 (2019).

26. Yan, M., Huang, X., Wu, J., Deng W., Lu, J. Y. & Liu, Z. Y. Antichirality Emergent in Type-II Weyl Phononic Crystals. Phys. Rev. Lett. **130**, 266304 (2023).

27. Ma, S. et al. Gauge field induced chiral zero mode in five-dimensional Yang monopole metamaterials. Phys. Rev. Lett. **130**, 243801 (2023).

28. Hashimoto, K., Kimura, T. & Wu, X. Prog. Boundary conditions of Weyl semimetals. Theor. Exp. Phys. **2017**, 053I01 (2017).

29. Xi, X. et al. Observation of chiral edge states in gapped nanomechanical graphene. Sci. Adv. **7**, eabe1398 (2021).

30. Wang, M. et al. Observation of boundary induced chiral anomaly bulk states and their transport properties. Nat. Commun. **13**, 5916 (2022).

**Methods**

**Numerical simulation**

The finite-element method solver of COMSOL Multiphysics software is used to perform the full-wave simulations. For the bulk structure of the Weyl metamaterial, we apply the periodic boundary conditions in all three spatial directions of the saddle-shaped unit cell. To calculate the top/bottom Fermi-arc surface states and CESs, we consider a $1 \times 1 \times 7$ supercell applying PEC boundary conditions along the $z$ direction and periodic boundary conditions along the $x(y)$ directions, respectively. The coper of the Weyl metamaterial waveguide is considered as the PEC, and the rest volumes are dielectric (dielectric constant is 2.6) and air.

**Experiment**

Our experimental sample consists of $50 \times 50 \times 7$ unit cells, which are fabricated via the printed circuit boards technique by etching 4 mm-thick dielectric laminates with double-sided, 0.035 mm-thick copper cladding. Each printed layer is paired with a 2 mm-thick dielectric blank layer. An air hole with a 2.6 mm radius at the centre of each unit cell allows the electromagnetic probe to be inserted to measure the fields within the Weyl metamaterial waveguide. In the measurements, the phase and amplitude of the field spectra at each frequency are collected by a vector network analyser. Two electric dipole antennas connect the vector network analyser, serving as the probe and the source. Several measurements are performed to obtain the top/bottom Fermi-arc surface state and CES, respectively. In these measurements, the sources are placed at the centre position of the whole sample. For the top and bottom Fermi-arc surface state measurements, the probes are vertically inserted into the top and bottom layers through the air holes of the Weyl metamaterial waveguide, respectively. For the CES measurements, the probe is vertically inserted into the middle layer of the waveguide. The field spectra are scanned point by point with a step of 6 mm in the *x-y* plane.

**Acknowledgements**

The work at Zhejiang University was sponsored by the Key Research and Development Program of the Ministry of Science and Technology under Grants 2022YFA1405200 (Y.Y.), No. 2022YFA1404704 (H.C.), 2022YFA1404902 (H.C.), and 2022YFA1404900 (Y.Y.), the National Natural Science Foundation of China (NNSFC) under Grants No. 62175215 (Y.Y.), and No. 61975176 (H.C.), the Key Research and Development Program of Zhejiang Province under Grant No.2022C01036 (H.C.), the Fundamental Research Funds for the Central Universities (2021FZZX001-19) (Y.Y.), and the Excellent Young Scientists Fund Program (Overseas) of China (Y.Y.).

**Author contributions**

Y.Y. initiated the idea. N.H., M.L., R.Z., J.M., Z.Y., and Y.Y. provided theoretical explanations. N.H. and Y.Y. performed the numerical simulations. N.H. and Y.Y. designed the experiment. N.H., F. C., and L.Z. fabricated samples. N.H., F.C., M.L., R.Z., and W.L. carried out the measurements. N.H., F.C., M.L., R.Z., W.L., Q.C., and Y.P. analysed data. N.H. and Y.Y. wrote the manuscript with input from F.C., M.L., R.Z., W.L., Q.C., L.Z., Y.P., J.M., Z.Y., and H.C. H.C. and Y.Y. supervised the project.

**Competing interests**

The authors declare no competing interests.

**Additional information**

Supplementary information is available for this paper.

**Correspondence and requests for materials** should be addressed to Hongsheng Chen or Yihao Yang.